# DKP Equation in a Rotating Frame with Magnetic Cosmic String Background

*H. Hassanabadi* [*1] *and M. Hosseinpour*[1]

[1]*Physics Department, Shahrood University, Shahrood, Iran P. O. Box: 3619995161-316, Shahrood, Iran*

[*]*Corresponding author, Tel.:+98 232 4222522; fax: +98 273 3395270*

Email: h.hasanabadi@shahroodut.ac.ir

**Abstract**

We study the non-inertial effects on the covariant DKP equation in the curved space-time of a cosmic string. The calculations are done in the presence of a magnetic vector potential in the Coulomb gauge and the analytical solutions for the modified spectrum of the system are reported.

## 1. Introduction

The non-inertial effects, which arise from a rotating frame, provide us with a better insight into our physical phenomena and extend our understanding of mathematical foundations of our theories [1-7]. In quantum mechanics, such effects are often ascribed to the coupling between the angular momentum and the angular velocity [8, 9]. Until now, a number of interesting papers have been released which generalize the ordinary quantum mechanics problems into non-inertial frames [10,11] including the Fermi-Walker reference frame [12]. The non-inertial effects incorporated with nonrelativistic and relativistic quantum wave equations are well addressed in Refs. [13–23]. A particular kind of space-time with nontrivial topology is the cosmic string space-time. The cosmic string is recognized as a topological defect arising from symmetry breaking phase transition in the early Universe [24,25]. The analysis of cosmic strings within the framework of quantum mechanics has been well reported in Refs. [2,3,26]. A cosmic string is a linear defect that changes the topology of the medium when viewed globally. The space-time around a cosmic string is locally flat but not globally. The theory of general relativity predicts that gravitation is manifested as curvature of space-time. This curvature is characterized by the Riemann tensor. There are connections between topological properties of the space and local physical laws. The nontrivial topology of space-time, as well as its curvature, leads to a number of interesting gravitational effects. For example, it has been known that the energy levels of an atom placed in a gravitational field will be shifted as a result of the interaction of the atom with space-time curvature. Therefore, we have to consider the topology of the space-time in order to describe completely the physics of system.

On the other hand, the Duffin-Kemmer-Petiau (DKP) equation has an outstanding position in relativistic quantum mechanics as it enables us consider both spin-zero and spin-one bosons in a unified basis, i.e. it is a counterpart to Klein-Gordon and Proca equations [27-29]. The DKP equation resembles the Dirac equation in structure, with the gamma matrices being replaced by beta matrices. A survey on related topics to DKP equation can be found in Refs. [30-43] where the equation has been considered for various physical interactions and in connection with nuclear, particle, and mathematical physics [38-43]. The influence of topological defect in the dynamics of spin-1/2 particles as the solutions of the Dirac equation have been studied in various curved backgrounds with different coordinate frames. However, the same problem involves bosons with spin-zero and spin-one particles via DKP formalism has not been considered. Therefore, we believe that this problem deserves to be studied.

In this work, we are going to consider non-inertial effects due to rotating frames within the framework of DKP equation in the curved space-time of a cosmic sting. We obtain the solutions of the equation in the presence of a magnetic vector potential in polar coordinates in the Coulomb gauge. The work is structured as follows: in Sec. 2, we present the covariant DKP equation in the curved space-time of a cosmic sting. In Sec. 3, we build the rotating frame and determine the field configuration in this rotating frame. Sec. 4 includes the solution of radial part of DKP Hamiltonian and the next section considers the special case of the ordinary space-time without rotating.

## 2. Duffin-Kemmer-Petiau Equation

DKP equation has been a promising framework to study relativistic scalar and vector bosons I various fields of physics including nuclear and particle physics [34,44,45]. The free-particle DKP Hamiltonian is

$$H_f^{DKP} = \vec{\alpha}\cdot\vec{p} + \beta^0 M \tag{1}$$

where $\alpha^i = \beta^0 \beta^i$. DKP equation can be equivalently written as [27,28]

$$(i\beta^\mu \partial_\mu - M)\Psi = 0 \tag{2}$$

where $M$ and $\Psi$ are the boson mass and the ten-component wave function, respectively, and $\beta^\mu$ denotes the DKP matrices which satisfy

$$\beta^\mu \beta^\nu \beta^\lambda + \beta^\lambda \beta^\nu \beta^\mu = g^{\mu\nu}\beta^\lambda + g^{\lambda\nu}\beta^\mu \tag{3}$$

with $g^{\mu\nu} = \text{diag}(1,-1,-1,-1)$ being defined as the metric tensor of Minkowski space-time.

## 3. DKP Equation in Fermi-Walker Reference Frame

In this section, we discuss the influence of the non-inertial effects of the Fermi-Walker reference frame on the DKP equation in the cosmic string space-time. We choose the cosmic string space-time background in an inertial frame where the line element of the Minkowski space-time background is given by [46,47]

$$ds^2 = -d\tau^2 + dR^2 + R^2\eta^2 d\Phi^2 + dZ^2 \tag{4}$$

Where $\eta = 1 - 4\tilde{m}$ is the parameter related to the angular deficit. It is defined in the range $0 < \eta < 1$, with $\tilde{m}$ being the linear mass density. In our calculations, we consider $\hbar = c = 1$ and the acceptable physical ranges of the parameters are $-\infty < z < \infty, \rho \geq 0$ and $0 \leq \varphi \leq 2\pi$. It should be noted that the deficit angle can assume only values $\eta < 1$. We build the Fermi-Walker reference frame by rotating the frame with a constant angular velocity and make the following coordinate transformation:

$$T = t, \quad R = \rho, \quad \Phi = \varphi + \omega t, \quad Z = z \tag{5}$$

where $\omega$ is the constant angular velocity of the rotating frame and must satisfy $\omega\rho \ll 1$. With this transformation, the line element (4) becomes [39-41].

$$ds^2 = -\left(1 - \omega^2\eta^2\rho^2\right)dt^2 + 2\omega\eta^2\rho^2 d\varphi dt + d\rho^2 + \eta^2\rho^2 d\varphi^2 + dz^2 \tag{6}$$

Where describes a scenario of general relativity corresponding to the cosmic string space-time background in a rotating coordinate system. We should note that the line element Eq.(6) is defined in the range $\rho < \frac{1}{\omega\eta}$. Values where $\rho > \frac{1}{\omega\eta}$ mean that the line element Eq.(6) is not well-defined because this region of the space-time corresponds to a particle placed outside of the light-cone because the velocity of the particle is greater than the velocity of the Light [11,12][23].

The covariant form of the DKP equation for non-interacting bosons is given by [27-29]

$$\left[i\beta^\mu \nabla_\mu - M\right]\Psi = 0 \tag{7}$$

where $\Psi$ is the boson wave function, $M$ denotes the mass, and $\beta^\mu$, Kemmer matrices, satisfy the commutation relation

$$\beta^\mu \beta^\nu \beta^\lambda + \beta^\lambda \beta^\nu \beta^\mu = g^{\mu\nu}\beta^\lambda + g^{\lambda\nu}\beta^\mu \tag{8}$$

The Kemmer matrices $\beta$ in curved space-time are related to their Minkowski counterparts via

$$\beta^\mu = e^\mu_{\bar{a}} \beta^{\bar{a}} \tag{9}$$

$$\beta^\mu = \gamma^\mu \otimes I + I \otimes \gamma^\mu \tag{10}$$

where $\gamma^\mu(x)$ are the generalized Dirac matrices which satisfy the anti-commutation relations

$$\{\gamma^\mu(x), \gamma^\nu(x)\} = 2g^{\mu\nu}(x) \tag{11}$$

The generalized Dirac matrices are written in terms of the standard Dirac matrices $\gamma^{\bar{a}}(x)$ in Minkowski space-time as

$$\gamma^{\mu} = e^{\mu}_{\bar{a}} \gamma^{\bar{a}} \tag{12}$$

With the line element given in the expression (6), we now intend to build a local reference frame to place the observers. It is in the local reference frame that we can define the spinor in the curved space-time background. We can build the local reference frame through a non-coordinate basis. Our building blocks for this aim are $e^{(a)}_{\mu}$ and $e^{\mu}_{(a)}(x)$ transformation matrices. The components of the non-coordinate basis $e^{(a)}_{\mu}$ are called tetrads or vierbeins that form our local reference frame and satisfy

$$\eta^{ab} e^{\mu}_{\bar{a}}(x) e^{\nu}_{\bar{b}}(x) = g^{\mu\nu}(x) \tag{13}$$

where $\mu, \nu = 0, 1, 2, 3$ are tensor indices and $\bar{a}, \bar{b} = 0, 1, 2, 3$ denote tetrad indices [48,49].

There are many ways to construct the tetrads. In order to write the DKP equation in this space-time we choose the tetrads and the inverses as [50]

$$e^{a}_{\mu}(x) = \begin{pmatrix} \sqrt{1-\beta^2} & 0 & -\dfrac{\omega \eta^2 \rho^2}{\sqrt{1-\beta^2}} & 0 \\ 0 & 1 & 0 & 0 \\ 0 & 0 & \dfrac{\eta\rho}{\sqrt{1-\beta^2}} & 0 \\ 0 & 0 & 0 & 1 \end{pmatrix},$$

$$e^{\mu}_{a}(x) = \begin{pmatrix} \dfrac{1}{\sqrt{1-\beta^2}} & 0 & \dfrac{\omega \eta \rho}{\sqrt{1-\beta^2}} & 0 \\ 0 & 1 & 0 & 0 \\ 0 & 0 & \dfrac{\sqrt{1-\beta^2}}{\eta\rho} & 0 \\ 0 & 0 & 0 & 1 \end{pmatrix} \tag{14}$$

Where $\beta = \omega\eta\rho$. The matrices $\gamma^{\mu}$ in Eq. (12) are given more explicitly as

$$\gamma^0 = e^{0}_{\bar{a}} \gamma^{\bar{a}} = e^{0}_{\bar{0}} \gamma^{\bar{0}} + e^{0}_{\bar{1}} \gamma^{\bar{1}} + e^{0}_{\bar{2}} \gamma^{\bar{2}} + e^{0}_{\bar{3}} \gamma^{\bar{3}} = \frac{1}{\sqrt{1-\beta^2}} \gamma^{\bar{0}} + \frac{\omega\eta\rho}{\sqrt{1-\beta^2}} \gamma^{\bar{2}} = \frac{\gamma^{\bar{0}} + \omega\eta\rho\gamma^{\bar{2}}}{\sqrt{1-\beta^2}}$$

$$\gamma^{t} = \gamma^{\bar{0}} + \omega\eta\rho\gamma^{\bar{2}} = \begin{pmatrix} 1 & \omega\eta\rho\sigma^2 \\ -\omega\eta\rho\sigma^2 & -1 \end{pmatrix} \tag{15-a}$$

$$\gamma^1 = e^{1}_{\bar{a}} \gamma^{\bar{a}} = e^{1}_{\bar{0}} \gamma^{\bar{0}} + e^{1}_{\bar{1}} \gamma^{\bar{1}} + e^{1}_{\bar{2}} \gamma^{\bar{2}} + e^{1}_{\bar{3}} \gamma^{\bar{3}} = \gamma^{\bar{1}} = \gamma^{\rho} \tag{15-b}$$

$$\gamma^2 = e^{2}_{\bar{a}} \gamma^{\bar{a}} = e^{2}_{\bar{0}} \gamma^{\bar{0}} + e^{2}_{\bar{1}} \gamma^{\bar{1}} + e^{2}_{\bar{2}} \gamma^{\bar{2}} + e^{2}_{\bar{3}} \gamma^{\bar{3}} = \frac{\sqrt{1-\beta^2}}{\eta\rho} \gamma^{\bar{2}} = \left( \frac{1}{\sqrt{1-\beta^2}\eta\rho} - \frac{\omega^2 \eta \rho}{\sqrt{1-\beta^2}} \right) \gamma^{\bar{2}} \tag{15-c}$$

$$\gamma^3 = e_0^3 \gamma^{\bar{0}} + e_1^3 \gamma^{\bar{1}} + e_2^3 \gamma^{\bar{2}} + e_3^3 \gamma^{\bar{3}} = \gamma^{\bar{3}} = \gamma^z \quad (15\text{-d})$$

As a result, the matrices $\beta^\mu$ in Eq. (10) can be written as

$$\beta^0 = e_0^0 \beta^{\bar{0}} + e_1^0 \beta^{\bar{1}} + e_2^0 \beta^{\bar{2}} + e_3^0 \beta^{\bar{3}} = \frac{1}{\sqrt{1-\beta^2}} \beta^{\bar{0}} + \frac{\omega\eta\rho}{\sqrt{1-\beta^2}} \beta^{\bar{2}} = \frac{\beta^{\bar{0}} + \omega\eta\rho\beta^{\bar{2}}}{\sqrt{1-\beta^2}} = \frac{\beta^t}{\sqrt{1-\beta^2}} \quad (16\text{-a})$$

$$\beta^1 = e_{\bar{a}}^1 \beta^{\bar{a}} = e_0^1 \beta^{\bar{0}} + e_1^1 \beta^{\bar{1}} + e_2^1 \beta^{\bar{2}} + e_3^1 \beta^{\bar{3}} = \beta^{\bar{1}} = \beta^\rho \quad (16\text{-b})$$

$$\beta^2 = e_{\bar{a}}^2 \beta^{\bar{a}} = e_0^2 \beta^{\bar{0}} + e_1^2 \beta^{\bar{1}} + e_2^2 \beta^{\bar{2}} + e_3^2 \beta^{\bar{3}} = \frac{\sqrt{1-\beta^2}}{\eta\rho} \beta^{\bar{2}} = \frac{\sqrt{1-\beta^2}\,\beta^{\bar{2}}}{\eta\rho} = \frac{\beta^\varphi}{\eta\rho} \quad (16\text{-c})$$

$$\beta^3 = e_0^3 \beta^{\bar{0}} + e_1^3 \beta^{\bar{1}} + e_2^3 \beta^{\bar{2}} + e_3^3 \beta^{\bar{3}} = \beta^{\bar{3}} = \beta^z \quad (16\text{-d})$$

The covariant derivative in Eq. (7) is

$$\nabla_\mu = \partial_\mu + \Omega_\mu$$

$$\Omega_\mu = \Gamma_\mu \otimes I + I \otimes \Gamma_\mu \quad (17)$$

and the affine connections are written as [51, 52]

$$\Gamma_\mu = \frac{i}{4} \omega_{\mu\bar{a}\bar{b}} \Sigma^{\bar{a}\bar{b}} \quad (18)$$

$$\Sigma^{\bar{a}\bar{b}} = \frac{i}{2} \left[ \gamma^{\bar{a}}, \gamma^{\bar{b}} \right]$$

w

*or*

$$\Gamma_\mu = \frac{-1}{4} \omega_\mu^{\bar{a}\bar{b}} \left[ \gamma^{\bar{a}}, \gamma^{\bar{b}} \right] \quad (19)$$

here $\gamma^{\bar{a}}$ are the standard Dirac matrices in Minkowski space-time and $\omega_{\mu\bar{a}\bar{b}}$ is the spin connection, given by

$$\omega_{\mu\bar{b}}^{\bar{a}} = e_\nu^{\bar{a}} e_{\bar{b}}^\sigma \Gamma_{\sigma\mu}^\nu + e_\nu^{\bar{a}} \partial_\mu e_{\bar{b}}^\nu \quad (20)$$

The non-vanishing components of the spin connection are [51,52]

$$\omega^0{}_{t\,1} = \omega^1{}_{t\,0} = -\frac{\omega\eta^2\rho}{\sqrt{1-\beta^2}}$$

$$\omega^1{}_{t\,2} = -\omega^2{}_{t\,1} = -\frac{\omega\eta}{\sqrt{1-\beta^2}}$$

$$\omega^0{}_{\rho\,2} = \omega^2{}_{\rho\,0} = \frac{\omega\eta}{1-\beta^2} \qquad (21)$$

$$\omega^0{}_{\varphi\,1} = \omega^1{}_{\varphi\,0} = -\frac{\omega\eta^2\rho}{\sqrt{1-\beta^2}}$$

$$\omega^1{}_{\varphi\,2} = -\omega^2{}_{\varphi\,1} = -\frac{\eta}{\sqrt{1-\beta^2}}$$

The non-vanishing spinor affine connection in Eq. (18) is written as

$$\Gamma_t = -\frac{1}{2}\frac{\omega^2\eta^2\rho}{\sqrt{1-\beta^2}}\hat{\alpha}^1 - \frac{i}{2}\frac{\omega\eta}{\sqrt{1-\beta^2}}\Sigma^3$$

$$\Gamma_\rho = \frac{1}{2}\frac{\omega\eta}{(1-\beta^2)}\hat{\alpha}^2 \qquad (22)$$

$$\Gamma_\varphi = -\frac{1}{2}\frac{\omega\eta^2\rho}{\sqrt{1-\beta^2}}\hat{\alpha}^1 - \frac{i}{2}\frac{\eta}{\sqrt{1-\beta^2}}\Sigma^3$$

where

$$\Sigma^3 = \begin{pmatrix} \sigma^3 & 0 \\ 0 & \sigma^3 \end{pmatrix} \qquad (23)$$

By substituting $\Gamma_\varphi$ in Eq. (17) we have [50]

$$\Omega_t = -\frac{1}{2}\frac{\omega^2\eta^2\rho}{\sqrt{1-\beta^2}}\begin{pmatrix} 0 & \sigma_1 & \sigma_1 & 0 \\ \sigma_1 & 0 & 0 & \sigma_1 \\ \sigma_1 & 0 & 0 & \sigma_1 \\ 0 & \sigma_1 & \sigma_1 & 0 \end{pmatrix} - \frac{i}{2}\frac{2\omega\eta}{\sqrt{1-\beta^2}}\begin{pmatrix} \sigma_3 & 0 & 0 & 0 \\ 0 & \sigma_3 & 0 & 0 \\ 0 & 0 & \sigma_3 & 0 \\ 0 & 0 & 0 & \sigma_3 \end{pmatrix}$$

$$\Omega_\rho = \frac{1}{2}\frac{\omega\eta}{(1-\beta^2)}\begin{pmatrix} 0 & \sigma_2 & \sigma_2 & 0 \\ \sigma_2 & 0 & 0 & \sigma_2 \\ \sigma_2 & 0 & 0 & \sigma_2 \\ 0 & \sigma_2 & \sigma_2 & 0 \end{pmatrix} \qquad (24)$$

$$\Omega_\varphi = -\frac{1}{2}\frac{\omega\eta^2\rho}{\sqrt{1-\beta^2}}\begin{pmatrix} 0 & \sigma_1 & \sigma_1 & 0 \\ \sigma_1 & 0 & 0 & \sigma_1 \\ \sigma_1 & 0 & 0 & \sigma_1 \\ 0 & \sigma_1 & \sigma_1 & 0 \end{pmatrix} - \frac{i}{2}\frac{2\eta}{\sqrt{1-\beta^2}}\begin{pmatrix} \sigma_3 & 0 & 0 & 0 \\ 0 & \sigma_3 & 0 & 0 \\ 0 & 0 & \sigma_3 & 0 \\ 0 & 0 & 0 & \sigma_3 \end{pmatrix}$$

The covariant DKP Eq. (7), written in the space-time of a cosmic string, is then given by [27-29]

$$\left[i\beta^0\partial_0 + i\beta^a\partial_a - M\right]\Psi = 0, (a = 1,2,3)$$
$$\left[i\left(I\otimes\gamma^\circ + \gamma^\circ\otimes I\right)\partial_0 + i\left(I\otimes\gamma^i + \gamma^i\otimes I\right)\partial_i - M\right]\Psi = 0 \quad (25)$$

By multiplying Eq. (25) from right by $\gamma^0\otimes\gamma^0$, we have

$$\left[i\left(\gamma^0\otimes\gamma^0\right)\left(I\otimes\gamma^\circ + \gamma^\circ\otimes I\right)\partial_0 + i\left(\gamma^0\otimes\gamma^0\right)\left(I\otimes\gamma^i + \gamma^i\otimes I\right)\nabla_i - \left(\gamma^0\otimes\gamma^0\right)M\right]\Psi = 0 \quad (26)$$

From $(A\otimes B)(C\otimes D) = AC\otimes BD$ we have

$$\left[i\left(\gamma^0\otimes I + I\otimes\gamma^0\right)\partial_0 + i\left(\gamma^0\otimes\alpha^i + \alpha^i\otimes\gamma^0\right)\nabla_i - \left(\gamma^0\otimes\gamma^0\right)M\right]\Psi = 0 \quad (27)$$

By substituting $\nabla_i$ form Eq. (17), Eq. (27) appears as [28]

$$\left[i\left(\gamma^0\otimes I + I\otimes\gamma^0\right)\left(\partial_\circ + \Omega_t\right) + i\left(\gamma^0\otimes\gamma^0\gamma^1 + \gamma^0\gamma^1\otimes\gamma^0\right)\left(\partial_1 + \Omega_\rho\right)\right.$$
$$\left. + i\left(\gamma^0\otimes\gamma^0\gamma^2 + \gamma^0\gamma^2\otimes\gamma^0\right)\left(\partial_2 + \Omega_\varphi\right) + i\left(\gamma^0\otimes\gamma^0\gamma^3 + \gamma^0\gamma^3\otimes\gamma^0\right)\left(\partial_3\right) - \left(\gamma^0\otimes\gamma^0\right)M\right]\Psi = 0 \quad (28)$$

Now, by substituting $\gamma^\circ, \gamma^1, \gamma^2, \gamma^3$ from Eq. (15), we write Eq. (28) in the form

$$\left\{i\left(\left(\gamma^\circ + \omega\eta\rho\gamma^{\bar{2}}\right)\otimes I + I\otimes\left(\gamma^\circ + \omega\eta\rho\gamma^{\bar{2}}\right)\right)\sqrt{1-\beta^2}\left(\partial_t + \Omega_t\right)\right.$$
$$+ i\left(\left(\gamma^\circ + \omega\eta\rho\gamma^{\bar{2}}\right)\otimes\left(\gamma^\circ + \omega\eta\rho\gamma^{\bar{2}}\right)\gamma^{\bar{1}} + \left(\gamma^\circ + \omega\eta\rho\gamma^{\bar{2}}\right)\gamma^{\bar{1}}\otimes\left(\gamma^\circ + \omega\eta\rho\gamma^{\bar{2}}\right)\right)\left(\partial_\rho + \Omega_\rho\right)$$
$$+ i\left(\left(\gamma^\circ + \omega\eta\rho\gamma^{\bar{2}}\right)\otimes\left(\gamma^\circ + \omega\eta\rho\gamma^{\bar{2}}\right)\gamma^{\bar{2}} + \left(\gamma^\circ + \omega\eta\rho\gamma^{\bar{2}}\right)\gamma^{\bar{2}}\otimes\left(\gamma^\circ + \omega\eta\rho\gamma^{\bar{2}}\right)\right)\left(\frac{\partial_\varphi + \Omega_\varphi}{\eta\rho}\right) \quad (29)$$
$$+ i\left(\left(\gamma^\circ + \omega\eta\rho\gamma^{\bar{2}}\right)\otimes\left(\gamma^\circ + \omega\eta\rho\gamma^{\bar{2}}\right)\gamma^{\bar{3}} + \left(\gamma^\circ + \omega\eta\rho\gamma^{\bar{2}}\right)\gamma^{\bar{3}}\otimes\left(\gamma^\circ + \omega\eta\rho\gamma^{\bar{2}}\right)\right)\left(\partial_z\right)$$
$$\left. - M\left(\left(\gamma^\circ + \omega\eta\rho\gamma^{\bar{2}}\right)\otimes\left(\gamma^\circ + \omega\eta\rho\gamma^{\bar{2}}\right)\right)\right\}\Psi = 0$$

The magnetic vector potential in polar coordinates in the Coulomb gauge is

$$e\vec{A} = \frac{B_\circ}{2}r\hat{\varphi} \quad (30)$$

We now introduce the generalized momentum as $\vec{\pi} = -i\left(\vec{\nabla} + \vec{\Omega}\right) - e\vec{A}$ where [53]

$$\vec{\nabla} = \frac{\partial}{\partial\rho}\hat{\rho} + \frac{1}{\eta\rho}\frac{\partial}{\partial\varphi}\hat{\varphi} + \frac{\partial}{\partial z}\hat{z} \quad (31)$$

is the gradient operator in polar coordinates. The covariant DKP Eq. (7), written in the space-time of a cosmic string is therefore given by

$$\left\{ i\left(\tilde{\gamma}^t \otimes I + I \otimes \tilde{\gamma}^t\right)\sqrt{1-\beta^2}\left(\partial_t + \Omega_t\right) + i\left(\tilde{\gamma}^t \otimes \tilde{\gamma}^t \gamma^{\bar{1}} + \tilde{\gamma}^t \gamma^{\bar{1}} \otimes \tilde{\gamma}^t\right)\left(\partial_\rho + \Omega_\rho\right) \right.$$

$$\left. + i\left(\tilde{\gamma}^t \otimes \tilde{\gamma}^t \gamma^{\bar{2}} + \tilde{\gamma}^t \gamma^{\bar{2}} \otimes \tilde{\gamma}^t\right)\left(\frac{\partial_\varphi + \Omega_\varphi}{\eta\rho}\right) + i\left(\tilde{\gamma}^t \otimes \tilde{\gamma}^t \gamma^{\bar{3}} + \tilde{\gamma}^t \gamma^{\bar{3}} \otimes \tilde{\gamma}^t\right)\left(\partial_z\right) - M\left(\tilde{\gamma}^t \otimes \tilde{\gamma}^t\right) \right\}\Psi = 0 \quad (32)$$

$$\left\{ i\left(\tilde{\gamma}^t \otimes I + I \otimes \tilde{\gamma}^t\right)\sqrt{1-\beta^2}\left(\partial_t + \Omega_t\right) + i\left(\tilde{\gamma}^t \otimes \vec{\tilde{\alpha}} + \vec{\tilde{\alpha}} \otimes \tilde{\gamma}^t\right)\cdot\vec{\pi} - M\left(\tilde{\gamma}^t \otimes \tilde{\gamma}^t\right) \right\}\Psi = 0 \quad (33)$$

$$\tilde{\alpha}^i = \tilde{\gamma}^t \gamma^i = \begin{pmatrix} -\beta\sigma^2\sigma^i & \sigma^i \\ -\sigma^i & -\beta\sigma^2\sigma^i \end{pmatrix}$$

$$\tilde{\gamma}^t \otimes \tilde{\alpha}^i + \tilde{\alpha}^i \otimes \tilde{\gamma}^t = \begin{pmatrix} -2\beta\sigma^2\sigma^i & \sigma^i - \beta^2\sigma^2\sigma^i\sigma^2 & (1-\beta^2)\sigma^i & \beta(\sigma^2\sigma^i + \sigma^i\sigma^2) \\ -\sigma^i + \beta^2\sigma^2\sigma^i\sigma^2 & 0 & -\beta(\sigma^2\sigma^i + \sigma^i\sigma^2) & -(1+\beta^2)\sigma^i \\ -(1-\beta^2)\sigma^i & -\beta(\sigma^2\sigma^i + \sigma^i\sigma^2) & 0 & -(\sigma^i + \beta^2\sigma^2\sigma^i\sigma^2) \\ \beta(\sigma^2\sigma^i + \sigma^i\sigma^2) & (1+\beta^2)\sigma^i & \sigma^i + \beta^2\sigma^2\sigma^i\sigma^2 & 2\beta\sigma^2\sigma^i \end{pmatrix} \quad (34)$$

$$\left\{ i\left(\tilde{\gamma}^t \otimes I + I \otimes \tilde{\gamma}^t\right)\sqrt{1-\beta^2}\left[\partial_t + -\frac{1}{2}\frac{\omega^2\eta^2\rho}{\sqrt{1-\beta^2}}\begin{pmatrix} 0 & \sigma_1 & \sigma_1 & 0 \\ \sigma_1 & 0 & 0 & \sigma_1 \\ \sigma_1 & 0 & 0 & \sigma_1 \\ 0 & \sigma_1 & \sigma_1 & 0 \end{pmatrix} - \frac{i}{2}\frac{2\omega\eta}{\sqrt{1-\beta^2}}\begin{pmatrix} \sigma_3 & 0 & 0 & 0 \\ 0 & \sigma_3 & 0 & 0 \\ 0 & 0 & \sigma_3 & 0 \\ 0 & 0 & 0 & \sigma_3 \end{pmatrix}\right] \right.$$

$$+ i\left(\tilde{\gamma}^t \otimes \tilde{\alpha}^1 + \tilde{\alpha}^1 \otimes \tilde{\gamma}^t\right)\cdot\left[\partial\rho + \frac{1}{2}\frac{\omega\eta}{(1-\beta^2)}\begin{pmatrix} 0 & \sigma_2 & \sigma_2 & 0 \\ \sigma_2 & 0 & 0 & \sigma_2 \\ \sigma_2 & 0 & 0 & \sigma_2 \\ 0 & \sigma_2 & \sigma_2 & 0 \end{pmatrix}\right]$$

$$+ i\left(\tilde{\gamma}^t \otimes \tilde{\alpha}^2 + \tilde{\alpha}^2 \otimes \tilde{\gamma}^t\right)\cdot\left[\frac{\partial\varphi}{\eta\rho} - \frac{1}{2}\frac{\omega\eta}{\sqrt{1-\beta^2}}\begin{pmatrix} 0 & \sigma_1 & \sigma_1 & 0 \\ \sigma_1 & 0 & 0 & \sigma_1 \\ \sigma_1 & 0 & 0 & \sigma_1 \\ 0 & \sigma_1 & \sigma_1 & 0 \end{pmatrix} - \frac{i}{\rho\sqrt{1-\beta^2}}\begin{pmatrix} \sigma_3 & 0 & 0 & 0 \\ 0 & \sigma_3 & 0 & 0 \\ 0 & 0 & \sigma_3 & 0 \\ 0 & 0 & 0 & \sigma_3 \end{pmatrix}\right]$$

$$\left. i\left(\tilde{\gamma}^t \otimes \tilde{\alpha}^3 + \tilde{\alpha}^3 \otimes \tilde{\gamma}^t\right)\cdot\left(\partial z\right) - M\left(\tilde{\gamma}^t \otimes \tilde{\gamma}^t\right) \right\}\begin{pmatrix} \Psi_1 \\ \Psi_2 \\ \Psi_3 \\ \Psi_4 \end{pmatrix} = 0 \quad (35)$$

The stationary state $\Psi$ in Eq. (35) is the four-component wave function of the Kemmer equation. Substituting this solution into Eq. (35) we have

$$\left\{ E\sqrt{1-\beta^2} \begin{pmatrix} 2 & \beta\sigma^2 & \beta\sigma^2 & 0 \\ -\beta\sigma^2 & 0 & 0 & \beta\sigma^2 \\ -\beta\sigma^2 & 0 & 0 & \beta\sigma^2 \\ 0 & -\beta\sigma^2 & -\beta\sigma^2 & 0 \end{pmatrix} - \frac{\omega^2\eta^2\rho i}{2} \begin{pmatrix} -2i\beta\sigma^3 & 2\sigma^1 & 2\sigma^1 & -2i\beta\sigma^3 \\ 0 & 0 & 0 & 0 \\ 0 & 0 & 0 & 0 \\ 2i\beta\sigma^3 & -2\sigma^1 & -2\sigma_1 & 2i\beta\sigma^3 \end{pmatrix} \right.$$

$$+\omega\eta \begin{pmatrix} 2\sigma^3 & i\beta\sigma^1 & i\beta\sigma^1 & 0 \\ -i\beta\sigma^1 & 0 & 0 & i\beta\sigma^1 \\ -i\beta\sigma^1 & 0 & 0 & i\beta\sigma^1 \\ 0 & -i\beta\sigma^1 & -i\beta\sigma^1 & 2\sigma^3 \end{pmatrix} + i\partial_\rho \begin{pmatrix} 2i\beta\sigma^3 & \sigma^1(1+\beta^2) & \sigma^1(1-\beta^2) & 0 \\ \sigma^1(1-\beta^2) & 0 & -2i\beta\sigma^3 & -\sigma^1(1+\beta^2) \\ \sigma^1(1+\beta^2) & 2i\beta\sigma^3 & 0 & -\sigma^1(1-\beta^2) \\ 0 & -\sigma^1(1-\beta^2) & -\sigma^1(1+\beta^2) & -2i\beta\sigma^3 \end{pmatrix}$$

$$+\frac{i}{2}\frac{\omega\eta}{(1-\beta^2)} \begin{pmatrix} 2i\sigma^3 & 2\beta\sigma^1 & 2\beta\sigma^1 & 2i\sigma^3 \\ -2\beta\sigma^1 & -2i\beta^2\sigma^3 & -2i\beta^2\sigma^3 & -2\beta\sigma^1 \\ 2\beta\sigma^1 & 2i\beta^2\sigma^3 & 2i\beta^2\sigma^3 & 2\beta\sigma^1 \\ -2i\sigma^3 & 2\beta\sigma^1 & -2\beta\sigma^1 & -2i\sigma^3 \end{pmatrix} + i\frac{\partial\varphi}{\eta\rho} \begin{pmatrix} -2\beta & \sigma^2(1-\beta^2) & \sigma^2(1-\beta^2) & 2\beta \\ \sigma^2(1+\beta^2) & 0 & 0 & -\sigma^2(1+\beta^2) \\ \sigma^2(1+\beta^2) & 0 & 0 & -\sigma^2(1+\beta^2) \\ -2\beta & -\sigma^2(1-\beta^2) & \sigma^2(1-\beta^2) & 2\beta \end{pmatrix}$$

$$-\frac{i}{2}\frac{\omega\eta}{\sqrt{1-\beta^2}} \begin{pmatrix} -2i\sigma^3(1-\beta^2) & 0 & 0 & -2i\sigma^3(1-\beta^2) \\ 0 & 0 & 0 & 0 \\ 0 & 0 & 0 & 0 \\ 2i\sigma^3(1-\beta^2) & 0 & 0 & 2i\sigma^3(1-\beta^2) \end{pmatrix} + \frac{1}{\rho\sqrt{1-\beta^2}} \begin{pmatrix} -2\beta\sigma^3 & i\sigma^1(1-\beta^2) & i\sigma^1(1-\beta^2) & 2\beta\sigma^3 \\ i\sigma^1(1-\beta^2) & 0 & 0 & -i\sigma^1(1-\beta^2) \\ i\sigma^1(1-\beta^2) & 0 & 0 & -i\sigma^1(1-\beta^2) \\ -2\beta\sigma^3 & -i\sigma^1(1-\beta^2) & -i\sigma^1(1-\beta^2) & 2\beta\sigma^3 \end{pmatrix}$$

$$\left. i\partial_z \begin{pmatrix} -2i\beta\sigma^1 & \sigma^3(1+\beta^2) & \sigma^3(1-\beta^2) & 0 \\ \sigma^3(1-\beta^2) & 0 & 2i\beta\sigma^1 & -\sigma^3(1+\beta^2) \\ \sigma^3(1+\beta^2) & -2i\beta\sigma^1 & 0 & -\sigma^3(1-\beta^2) \\ 0 & -\sigma^3(1-\beta^2) & -\sigma^3(1+\beta^2) & 2i\beta\sigma^1 \end{pmatrix} - M \begin{pmatrix} 1 & \beta\sigma^2 & \beta\sigma^2 & \beta^2 \\ -\beta\sigma^2 & -1 & -\beta^2 & -\beta\sigma^2 \\ -\beta\sigma^2 & -\beta^2 & -1 & -\beta\sigma^2 \\ \beta^2 & \beta\sigma^2 & \beta\sigma^2 & 1 \end{pmatrix} \right\} \begin{pmatrix} \Psi_1 \\ \Psi_2 \\ \Psi_3 \\ \Psi_4 \end{pmatrix} = 0 \quad (36)$$

Substituting this solution into Eq. (36) leads to the four linear algebraic equations

$$\begin{aligned} A_1\Psi_1 + B_1\Psi_2 + C_1\Psi_3 + D_1\Psi_4 &= 0 \\ A_2\Psi_1 + B_2\Psi_2 + C_2\Psi_3 + D_2\Psi_4 &= 0 \\ A_3\Psi_1 + B_3\Psi_2 + C_3\Psi_3 + D_3\Psi_4 &= 0 \\ A_4\Psi_1 + B_4\Psi_2 + C_4\Psi_3 + D_4\Psi_4 &= 0 \end{aligned} \quad (37)$$

Where the coefficients of above equations are

$$A_1 = 2E\sqrt{1-\beta^2} - \omega^2\eta^2\rho\beta\sigma^3 + 2\omega\eta\sigma^3 - 2\beta\sigma^3\partial_\rho - \frac{\omega\eta\sigma^3}{(1-\beta^2)} - 2i\beta\frac{\partial\varphi}{\eta\rho} - \frac{\omega\eta}{\sqrt{1-\beta^2}}\sigma^3(1-\beta^2) - \frac{2\beta\sigma^3}{\rho\sqrt{1-\beta^2}} + 2\beta\sigma^1\partial_z - M$$

$$B_1 = E\sqrt{1-\beta^2}\sigma^2 - \omega^2\eta^2\rho i\sigma^1 + i\omega\eta\beta\sigma^1 + i\sigma^1(1+\beta^2)\partial_\rho + \frac{i\omega\eta\beta}{(1-\beta^2)}\sigma^1 + i\sigma^2\frac{(1-\beta^2)}{\eta\rho}\partial\varphi + i\sigma^1\frac{\sqrt{1-\beta^2}}{\rho} + i(1+\beta^2)\sigma^3\partial_z - M\beta\sigma^2$$

$$C_1 = E\sqrt{1-\beta^2}\beta\sigma^2 - \omega^2\eta^2\rho i\sigma^1 + i\omega\eta\beta\sigma^1 + i\sigma^1(1-\beta^2)\partial_\rho + \frac{i\omega\eta\beta}{(1-\beta^2)}\sigma^1 + i\sigma^2\frac{(1-\beta^2)}{\eta\rho}\partial\varphi + i\sigma^1\frac{\sqrt{1-\beta^2}}{\rho} + i(1-\beta^2)\sigma^3\partial_z - M\beta\sigma^2$$

$$D_1 = -\omega^2\eta^2\rho\beta\sigma^3 - \frac{\omega\eta}{(1-\beta^2)}\sigma^3 + 2i\beta\frac{\partial\varphi}{\eta\rho} - \omega\eta\sigma^3\sqrt{1-\beta^2} + \frac{2\beta\sigma^3}{\rho\sqrt{1-\beta^2}} - M\beta^2$$

(38-a)

$$A_2 = -E\sqrt{1-\beta^2}\beta\sigma^2 - i\omega\eta\beta\sigma^1 + i\sigma^1(1-\beta^2)\partial_\rho - \frac{i\omega\eta}{(1-\beta^2)}\sigma^1 + i\sigma^2\frac{(1+\beta^2)}{\eta\rho}\partial_\varphi + i\sigma^1\frac{\sqrt{1-\beta^2}}{\rho} + i(1-\beta^2)\sigma^3\partial_z + M\beta\sigma^2$$

$$B_2 = \frac{\omega\eta\sigma^3\beta^2}{1-\beta^2} + M, \qquad C_2 = 2\beta\sigma^3\partial_\rho + \frac{\omega\eta}{(1-\beta^2)}\sigma^3\beta^2 - 2\beta\sigma^1\partial_z + M\beta^2$$

$$D_2 = E\sqrt{1-\beta^2}\beta\sigma^2 + i\omega\eta\beta\sigma^1 - i\sigma^1(1+\beta^2)\partial_\rho - \frac{i\omega\eta\beta}{(1-\beta^2)}\sigma^1 - i\sigma^2\frac{(1+\beta^2)}{\eta\rho}\partial_\varphi - i\sigma^1\frac{\sqrt{1-\beta^2}}{\rho} - i(1-\beta^2)\sigma^3\partial_z + M\beta\sigma^2$$

(38-b)

$$A_3 = -E\sqrt{1-\beta^2}\beta\sigma^2 - i\omega\eta\beta\sigma^1 + i\sigma^1(1+\beta^2)\partial_\rho + \frac{\omega\eta}{(1-\beta^2)}\beta\sigma^1 + i\sigma^2\frac{(1+\beta^2)}{\eta\rho}\partial_\varphi + i\sigma^1\frac{(1+\beta^2)}{\rho\sqrt{1-\beta^2}} + i(1+\beta^2)\sigma^3\partial_z + M\beta\sigma^2$$

$$B_3 = -2\beta\sigma^3\partial_\rho - \frac{\omega\eta}{(1-\beta^2)}\sigma^3\beta^2 + 2\beta\sigma^1\partial_z + M\beta^2, \qquad C_3 = \frac{-\omega\eta\sigma^3\beta^2}{1-\beta^2} + M$$

$$D_3 = E\sqrt{1-\beta^2}\beta\sigma^2 + i\omega\eta\beta\sigma^1 - i\sigma^1(1-\beta^2)\partial_\rho + \frac{i\omega\eta\beta}{(1-\beta^2)}\sigma^1 - i\sigma^2\frac{(1+\beta^2)}{\eta\rho}\partial_\varphi - i\sigma^1\frac{(1+\beta^2)}{\rho\sqrt{1-\beta^2}} - i(1-\beta^2)\sigma^3\partial_z + M\beta\sigma^2$$

(38-c)

$$A_4 = \omega^2\eta^2\rho\beta\sigma^3 + \frac{\omega\eta}{(1-\beta^2)}\sigma^3 - 2i\beta\frac{\partial\varphi}{\eta\rho} + \omega\eta\sigma^3\sqrt{1-\beta^2} - \frac{2\beta\sigma^3}{\rho\sqrt{1-\beta^2}} - M\beta^2$$

$$B_4 = -E\sqrt{1-\beta^2}\beta\sigma^2 + \omega^2\eta^2\rho i\sigma^1 - i\omega\eta\beta\sigma^1 - i\sigma^1(1-\beta^2)\partial_\rho - \frac{i\omega\eta\beta}{(1-\beta^2)}\sigma^1 - i\sigma^2\frac{(1-\beta^2)}{\eta\rho}\partial_\varphi - i\sigma^1\frac{\sqrt{1-\beta^2}}{\rho} - i(1-\beta^2)\sigma^3\partial_z - M\beta\sigma^2$$

$$C_4 = -E\sqrt{1-\beta^2}\beta\sigma^2 + \omega^2\eta^2\rho i\sigma^1 - i\omega\eta\beta\sigma^1 - i\sigma^1(1+\beta^2)\partial_\rho - \frac{i\omega\eta\beta}{(1-\beta^2)}\sigma^1 - i\sigma^2\frac{(1-\beta^2)}{\eta\rho}\partial_\varphi - i\sigma^1\frac{\sqrt{1-\beta^2}}{\rho} - i(1+\beta^2)\sigma^3\partial_z - M\beta\sigma^2$$

$$D_4 = -2E\sqrt{1-\beta^2} + \omega^2\eta^2\rho\beta\sigma^3 - 2\omega\eta\sigma^3 + 2\beta\sigma^3\partial_\rho + \frac{\omega\eta\sigma^3}{(1-\beta^2)} + 2i\beta\frac{\partial\varphi}{\eta\rho} + \frac{\omega\eta}{\sqrt{1-\beta^2}}\sigma^3(1-\beta^2) + \frac{2\beta\sigma^3}{\rho\sqrt{1-\beta^2}} - 2\beta\sigma^1\partial_z - M$$

(38-d)

## 4. Special Case of $\omega \to 0$

The presence of localized curvature can have effects on geodesic motion and parallel transport in regions where the curvature vanishes. The best known example of this nonlocal effect is provided when a particle is transported around an idealized cosmic string along a closed curve in which case the string is noticed at all. Its gravitational counterpart may be viewed as a manifestation of nontrivial topology of space-time. The local intrinsic geometry of the space is not sufficient to describe completely the physics of a given system. In the following, we are going to consider the inertial condition by investigating the limit $\omega \to 0$. In this limit, the metric of Eq. (6) change to

$$ds^2 = -dt^2 + d\rho^2 + \eta^2\rho^2 d\varphi^2 + dz^2 \tag{39}$$

The nonvanishing component of $\vec{\Omega}$ is $\Omega_\varphi = -i\eta\Sigma^3\hat{\varphi}$. The generalized momentum has the form

$$\vec{\pi} = -i\left(\partial_\rho\hat{\rho} + \frac{1}{\eta\rho}(\partial_\varphi - i\eta\Sigma^3)\hat{\varphi} + \partial_z\hat{z}\right) = -i\left(\vec{\nabla} - i\frac{\eta\sigma^3}{\rho}\hat{\varphi}\right)$$

(40)

In the presence of electromagnetic potential, we have

$$\vec{\pi} = -i\left(\vec{\nabla} - i\frac{\eta\sigma^3}{\rho}\hat{\varphi}\right) - e\vec{A} = -i\vec{\nabla} - \frac{\eta\sigma^3}{\rho}\hat{\varphi} - \frac{B_\circ}{2}r\hat{\varphi} \qquad (41)$$

and the four linear algebraic Eqs.(37) yield

$$(2E - M)\Psi_1 + \left(-\vec{\sigma}.\vec{p} + i\frac{\sigma^1}{\rho}\right)\Psi_2 + \left(-\vec{\sigma}.\vec{p} + i\frac{\sigma^1}{\rho}\right)\Psi_3 = 0 \qquad \text{(42-a)}$$

$$\left(-\vec{\sigma}.\vec{p} + i\frac{\sigma^1}{\rho}\right)\Psi_1 + M\Psi_2 + \left(\vec{\sigma}.\vec{p} - i\frac{\sigma^1}{\rho}\right)\Psi_4 = 0 \qquad \text{(42-b)}$$

$$\left(-\vec{\sigma}.\vec{p} + i\frac{\sigma^1}{\rho}\right)\Psi_1 + M\Psi_3 + \left(\vec{\sigma}.\vec{p} - i\frac{\sigma^1}{\rho}\right)\Psi_4 = 0 \qquad \text{(42-c)}$$

$$\left(\vec{\sigma}.\vec{p} - i\frac{\sigma^1}{\rho}\right)\Psi_2 + \left(\vec{\sigma}.\vec{p} - i\frac{\sigma^1}{\rho}\right)\Psi_3 - (2E + M)\Psi_4 = 0 \qquad \text{(42-d)}$$

Now, from Eq. (41), we obtain

$$\vec{\sigma}.\vec{\pi} = -i\left(\sigma^1\partial_\rho + \frac{\sigma^2}{\eta\rho}(\partial_\varphi - i\eta\sigma^3)\hat{\varphi} + \sigma^3\partial_z\hat{z}\right) = -i\sigma^1\partial_\rho - i\frac{\sigma^2}{\eta\rho}\partial_\varphi - i\frac{\sigma^1}{\rho} - i\sigma^3\partial_z \qquad (43)$$

and rewrite Eqs. (42) as

$$(2E - M)\Psi_1 - \vec{\sigma}.\vec{\pi}\Psi_2 - \vec{\sigma}.\vec{\pi}\Psi_3 = 0 \qquad \text{(44-a)}$$

$$-\vec{\sigma}.\vec{\pi}\Psi_1 + M\Psi_2 + \vec{\sigma}.\vec{\pi}\Psi_4 = 0 \qquad \text{(44-b)}$$

$$-\vec{\sigma}.\vec{\pi}\Psi_1 + M\Psi_3 + \vec{\sigma}.\vec{\pi}\Psi_4 = 0 \qquad \text{(44-c)}$$

$$\vec{\sigma}.\vec{\pi}\Psi_2 + \vec{\sigma}.\vec{\pi}\Psi_3 - (2E + M)\Psi_4 = 0 \qquad \text{(44-d)}$$

By solving the above system of equations for $\Psi_2$, we have

$$\Psi_1 = \frac{2\vec{\sigma}.\vec{\pi}(\Psi_2)}{(2E - M)}, \qquad \Psi_2 = \Psi_3, \qquad \Psi_4 = \frac{2\vec{\sigma}.\vec{\pi}(\Psi_2)}{(2E + M)} \qquad (45)$$

and

$$\left\{(\vec{\sigma}.\vec{\pi}).(\vec{\sigma}.\vec{\pi}) + E^2 - \frac{M^2}{4}\right\}\Psi_2 = 0 \tag{46}$$

or

$$\left\{(\vec{\sigma}.\vec{\pi}).(\vec{\sigma}.\vec{\pi}) + E^2 - \frac{M^2}{4}\right\}\Psi_2 =$$

$$\left(\pi^2 + i\vec{\sigma}.\vec{\pi}\times\vec{\pi} + E^2 - \frac{M^2}{4}\right)\Psi_2 = \tag{46}$$

$$\left(-\nabla^2 - 2\frac{\sigma^3}{\eta\rho^2}m - \sigma^3 B_\circ + E^2 - \frac{M^2}{4}\right)\Psi_2 = 0$$

Where $\nabla^2$ is the Laplacian operator in the cosmic string background. Writing the explicit form of the Laplacian, we have

$$\left(-\frac{\partial^2}{\partial\rho^2} - \frac{1}{\rho}\frac{\partial}{\partial\rho} + \frac{1}{\rho^2}\left(\sigma^3 - \frac{m}{\eta}\right)^2 + \frac{B_\circ^2}{4}\rho^2 + B_\circ\left(\sigma^3 - \frac{m}{\eta}\right) - B_\circ\sigma^3 + \frac{M^2}{4} - E^2\right)\psi_2 = 0 \tag{47}$$

which gives the energy levels of the relativistic oscillator in the cosmic string space-time as

$$n = \frac{k_3}{4\sqrt{k_2}} - \frac{\sqrt{k_1}+1}{2} \tag{48}$$

Where

$$K_1 = \left(\sigma^3 - \frac{m}{\eta}\right)^2$$

$$K_2 = \frac{B_\circ^2}{4} \tag{49}$$

$$K_3 = B_\circ\left(\sigma^3 - \frac{m}{\eta}\right) - B_\circ\sigma^3 + \frac{M^2}{4} - E^2$$

As the final step, it should be mentioned that the corresponding wave function is

$$\Psi_2 = Nr^{\sqrt{k_1}} e^{-\frac{r^2}{2}\sqrt{k_2}} L_n^{\frac{\sqrt{k_1}}{2}-1}\left(r^2\sqrt{k_2}\right) \tag{50}$$

Where $N$ is the normalization constant and $L_n^{\frac{\sqrt{k_1}}{2}-1}$ denotes the Laguerre polynomials.

## Conclusions

We considered the covariant DKP equation in the curved space-time of a cosmic string and studied the non-inertial effects due to rotating frames. We calculated the solutions of the DKP equation in presence of a magnetic vector potential in polar coordinates in the Coulomb gauge and reported the analytical solutions.